\documentclass{article}

\usepackage{arxiv}
\usepackage{gensymb}
\usepackage[utf8]{inputenc} 
\usepackage[T1]{fontenc}    
\usepackage{hyperref}       
\usepackage{url}            
\usepackage{booktabs}       
\usepackage{amsfonts}       
\usepackage{nicefrac}       
\usepackage{microtype}      
\usepackage{lipsum}
\usepackage{graphicx}
\usepackage{subcaption}
\vspace{-2em}

\title{ Solar Radio Observation Using CALLISTO at the USO/PRL, Udaipur }

\author{
 Kushagra Upadhyay\\
 Udaipur Solar Observatory\\
  Physical Research Laboratory\\
   Udaipur 313 001, Rajasthan, India\\
  \texttt{kushagra@prl.res.in} \\
  \And
  Bhuwan Joshi\\
  Udaipur Solar Observatory\\
  Physical Research Laboratory\\
   Udaipur 313 001, Rajasthan, India\\
  \And
   Prabir K. Mitra\\
   Udaipur Solar Observatory\\
  Physical Research Laboratory\\
   Udaipur 313 001, Rajasthan, India\\
  \And
  Ramit Bhattacharyya\\
  Udaipur Solar Observatory\\
  Physical Research Laboratory\\
   Udaipur 313 001, Rajasthan, India\\
   \And
 Divya Oberoi \\
 National Centre for Radio Astrophysics \\
  Tata Institute of Fundamental Research\\
 Pune 411 007, Maharashtra, India \\
  \And
 Christian Monstein \\
  Istituto Ricerche Solari Locarno (IRSOL)\\
  Via Patocchi 57, 6605 Locarno Monti, Switzerland\\}

\date{}

\begin{document}

\maketitle
\begin{abstract}
This paper presents a detailed description of
various subsystems of CALLISTO solar radio
spectrograph installed at the USO/PRL. In the
front-end system, a log periodic dipole antenna (LPDA)
is designed for the frequency range of 40-900 MHz. In
this paper LPDA design, its modifications, and
simulation results are presented. We also present some
initial observations taken by CALLISTO at Udaipur. 
\end{abstract}

\section{Introduction}
e-CALLISTO is a worldwide network of radio
spectrometers. It is used for the observation of solar
radio bursts and radio frequency interference
monitoring for astronomical sciences. e-CALLISTO
system is a valuable tool for monitoring solar
activity and space weather research. Abbreviation
e-CALLISTO stands for \textbf{e}xtended \textbf{C}ompound
\textbf{A}stronomical \textbf{L}ow cost \textbf{L}ow frequency \textbf{I}nstrument
for \textbf{S}pectroscopy and \textbf{T}ransportable \textbf{O}bservatory. 
\par
The entire system of CALLISTO solar radio
spectrograph has successfully installed at Udaipur
Solar Observatory, Physical Research Laboratory,
Udaipur, India ( 24\degree 34'16.5720'' N and 73\degree 41'29.5584'' E) and has been observing the Sun
continuously since installation on October 3, 2018. A
proper data pipeline has been developed to process
data and transfer it to the server automatically.
\par
\begin{figure}[h!] 
    \centering
    \includegraphics[width=0.95\textwidth,height=\textheight,scale=1,keepaspectratio]{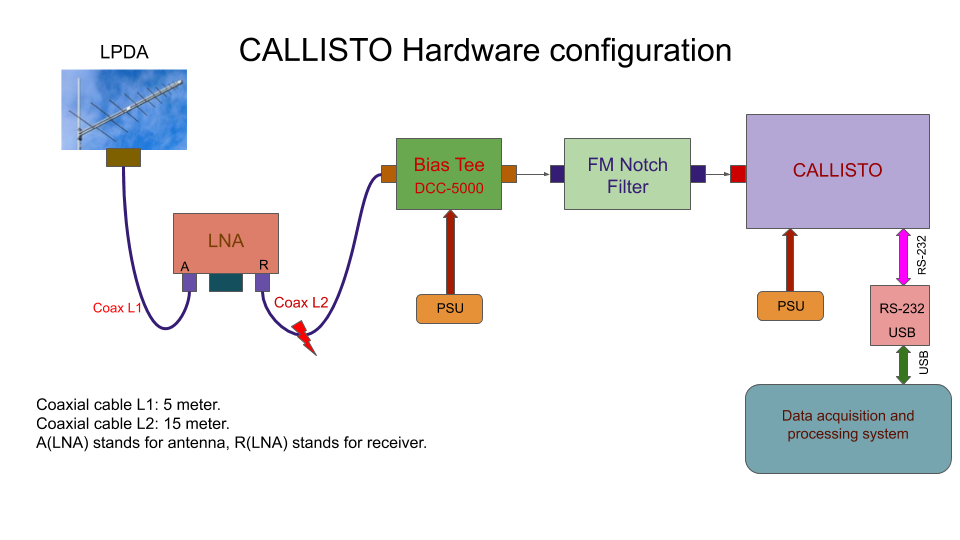}
    \label{Fig. 2}
    \caption{ Block diagram showing various subsystems of CALLISTO and their configuration}
\end{figure}
\par

\begin{figure}[h!]

    \centering
    \includegraphics[width=0.7\textwidth,height=\textheight,keepaspectratio]{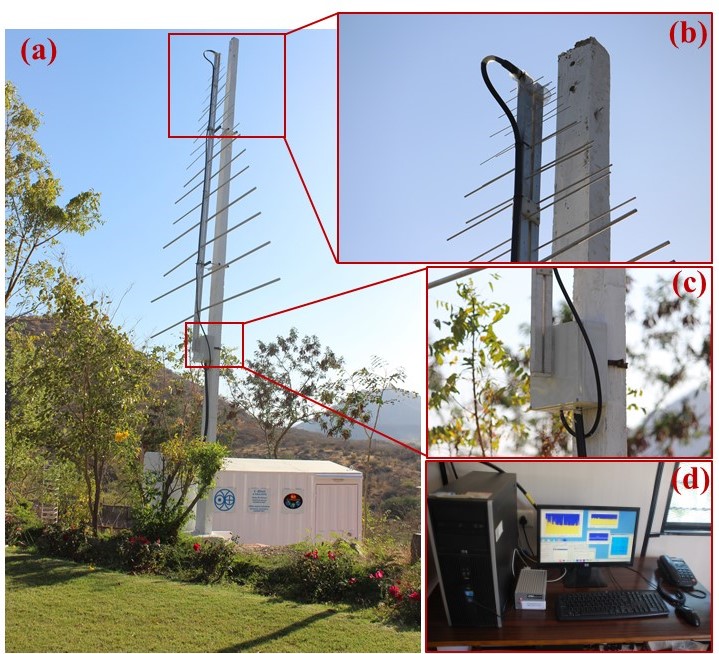}
    \label{Fig. 1}
    \caption{ Various subsystems of CALLISTO at the
USO/PRL are shown. Panel (a) shows the LPDA.
LNA (panel c) is connected between LPDA and
CALLISTO spectrograph (panel d). Panel (b) shows
zoomed view for thin elements of LPDA, which are
resonating at higher frequencies. }
\end{figure}

The essential component of the front-end part consists of a Log Periodic Dipole Antenna (LPDA).  It was designed  and simulated in CST software, and fabricated in NCRA, Pune workshop. Other front-end components, such as the low noise amplifier (LNA), FM rejection filter, bias tee, coaxial cables, and  connectors with very low attenuation losses were integrated with proper  impedance matching. The back-end  consists of e-CALLISTO spectrometer, which is a programmable heterodyne receiver designed at ETH Zurich, Switzerland.

\begin{figure}[!t] 
    \centering
    \includegraphics[width=\textwidth,height=\textheight,scale=1,keepaspectratio]{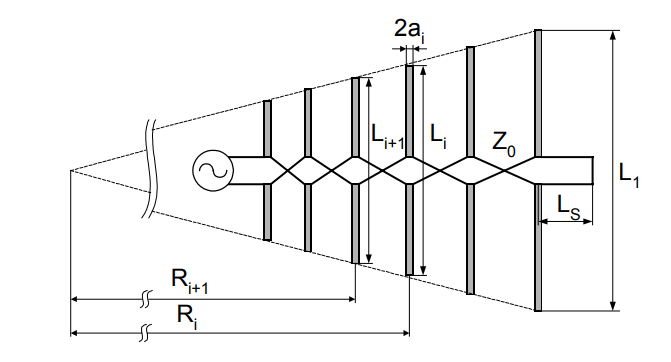}
    \label{Fig. 3}
    \caption{ Schematic of the log periodic dipole antenna (LPDA).}
\end{figure}

\begin{table}[!ht]
 \caption{Mechanical parameters of LPDA}
  \centering
  \begin{tabular}{ll}
    \toprule

    Number of elements &  28   \\
   Material for boom and elements & Aluminium      \\
    Length of each boom        & 3.63 m  \\
    Spacing between parallel boom  & 1 cm   \\
    Cross section  & 4 cm $\times$ 4 cm  \\
    Length of first dipole  &  3.75 m   \\
    Spacing between the first two elements &  0.45 m   \\
    Total stub length & 0.937 m \\
    \bottomrule
  \end{tabular}
  \label{tab:table}
\end{table}
\begin{table}[!ht]
 \caption{Diameter variations in the elements of LPDA}
  \centering
  \begin{tabular}{ll}
    \toprule

    Elements    &  Diameter    \\
    \midrule
  $  L_1  -  L_4   $ &  $\phi \hspace{1mm} 25 \hspace{1mm} mm $  \\
  $  L_5  -  L_7 $ & $ \phi \hspace{1mm}  19.5 \hspace{1mm} mm $     \\
   $  L_8  -  L_{\mathrm{12}} $  &  $ \phi \hspace{1mm}  9.5 \hspace{1mm} mm $  \\
  $  L_{\mathrm{13}}  -  L_{\mathrm{18}} $  & $ \phi  \hspace{1mm} 5 \hspace{1mm} mm $  \\
  $   L_{\mathrm{19}}  -  L_{\mathrm{28}} $ & $ \phi \hspace{1mm}  2 \hspace{1mm} mm $  \\
    \bottomrule
  \end{tabular}
  \label{tab:table}
\end{table}

\section{Log Periodic Dipole Antenna (LPDA)}

\subsection{Design and Mechanical structure}
Log periodic dipole antenna is a directional antenna that is used for wide band application at VHF/UHF. It is a multi element array in which the elements are arranged in a criss cross fashion to achieve phase reversal between consecutive elements. The mechanical phase reversal between these elements produces a phase progression so that the energy is beamed endfire in the direction of shorter elements. The most active elements in the feed  are those that are near resonant with a combined radiation pattern toward the vertex of the array [1].
\par
LPDA is designed using design equations given in [2]. Some basic equations are given below:
\par
$L_{\mathrm{i+1}}$= $\tau L_i$	…(1)
\par
$R_{\mathrm{i+1}}$= $\tau R_i$	…(2)
\par
$\sigma$ = $\frac{R_i-R_{\mathrm{i-1}}}{2L_{\mathrm{i-1}}}$	…(3)
\par
where $L_i$ is the length of ith dipole, $R_i$ is the spacing between with and (i+1)\textsubscript{th} element, $L_s$  is the stub length, $\tau$ is taper, and $\sigma$ is relative spacing (see Fig. 3).  
\par

 The value of taper ($\tau$) and relative spacing ($\sigma$) are finalized such that it provides a gain of 8 dBi and Voltage Standing Wave Ratio (VSWR) less than 2  in optimum length of antenna. Several simulations are done in CST with different values of $\tau$ and $\sigma$ and finally we arrived at optimal values of 0.88 and 0.06 respectively. Stub is used to improve VSWR performance and the short circuit stub length is taken as $\lambda$/8, where  is wavelength corresponding to the lowest frequency.
 Mechanical parameters for the fabrication of LPDA are given in Table 1.

\par
In the simulation, we found the diameter of the elements changes VSWR because impedance of dipole depend on length to diameter ratio (L/D); so, we tried to make L/D ratio relatively the same for the entire length of antenna. However,  practically there is a limitation regarding availability of different diameter of aluminium pipe in the market. Therefore, we have chosen 5 different diameter variants according to the availability (see Table 2) and mechanical design simulated in CST. 

\par
\subsection{ Simulation Results}

The  gain (G) of an antenna is given by the product of total efficiency ($\eta$) and directivity (D). From Fig. 4, we find that gain is approximately 8 dBi and H-plane Half Power Beam Width (HPBW) is approximately $90\degree$ in the entire band. Fig. 5 shows that the VSWR in the entire band (40-900 MHz) is below 2, which implies that the  return loss of the antenna is lower than -10 dB  in the entire band. 
 \par
 LPDA has a wider beam width so it does not necessarily require tracking system, and hence, offers less mechanical complexity although its performance increases with tracking system. We have fixed the LPDA with vertical mounting where the dipoles are in North - South direction. Our antenna has  HPBW of $90\degree$, which provides about $\pm3$ hours of good observations in the East-West direction from local noon (for details, see Fig. 5).
\section{ Low Noise Amplifier (LNA)}
We used broadband low noise amplifier having noise figure of 1.1 dB with input and output VSWR below 2.  It provides a gain of 18 - 20 dBi.
\section{FM Rejection Filter}
The RFI level at CALLISTO/Udaipur site is low but strong FM signals cause severe cross modulation of the real data in nearby frequencies. To prevent this, we have used a notch filter.
\section{CALLISTO Spectrograph}
It is a compact frequency agile spectrograph operating at the bandwidth of 45-870 MHz. It measures 200 frequency channels with a temporal resolution of 0.25 sec. It has a sensitivity of 25 mV/dB and dynamic range of -120 dBm to -20 dBm [3].

\section{Observations taken by  CALLISTO at the USO   }

\par
From Fig. 6, we identify few pre-existing standard frequencies dedicated for commercial and communication purposes. We have used those frequencies as a reference to calibrate the system performance. We used an FM Rejection filter which causes a trench like structure in 88-108 MHz band of the spectral overview. The scientific data is made available every 15 minutes as standard FITS file. Each FITS file  contains observation taken by the CALLISTO spectrograph during an interval of 15 minutes. A proper data pipeline has been developed that transfers the data to the server in real time. The JPEG images, constructed from FITS files, are also available in the webpage for the purpose of quick viewing. Further, a light curve is generated in 5 different frequencies which gets updated every 15 minutes on the webpage. For more details, refer to the following link: https://www.prl.res.in/~ecallisto/.
\begin{figure} 

    \centering
     \begin{subfigure}[t]{0.03\textwidth}
    \textbf{A}
  \end{subfigure}
 \begin{subfigure}{0.95\textwidth}
 \centering
      \includegraphics[width=0.8\linewidth,height=\textheight,scale=1,keepaspectratio]{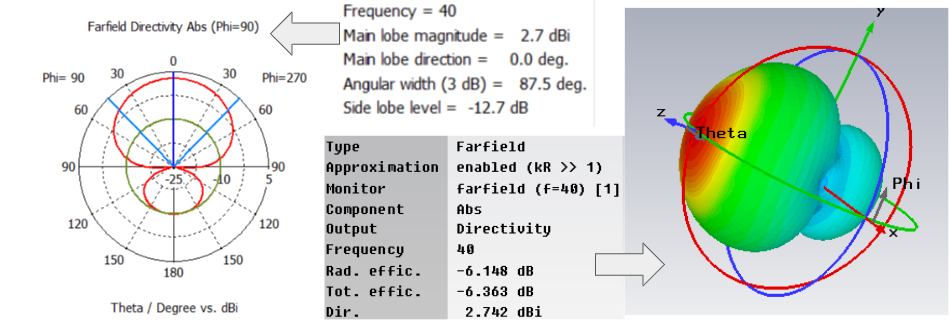}
       \end{subfigure}
 \begin{subfigure}{0.03\textwidth}
    \textbf{B}
  \end{subfigure}
   \begin{subfigure}{0.95\textwidth}
   \centering
    \includegraphics[width=0.8\linewidth,height=\textheight,scale=1,keepaspectratio]{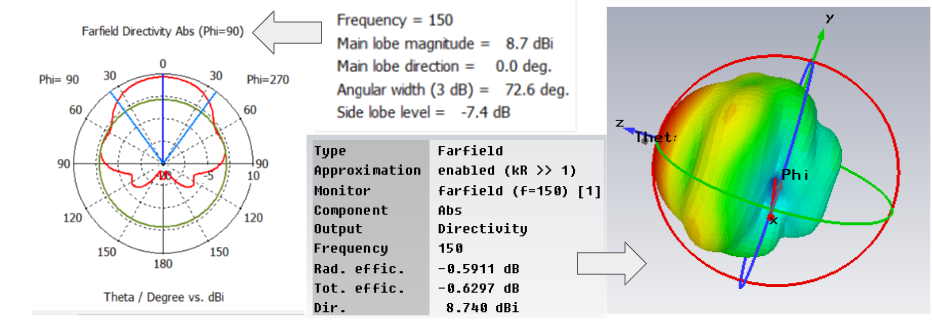}
    \end{subfigure}
    \begin{subfigure}{0.03\textwidth}
    \textbf{C}
    \end{subfigure}
    \begin{subfigure}{0.95\textwidth}
    \centering
     \includegraphics[width=0.8\linewidth,height=\textheight,scale=1,keepaspectratio]{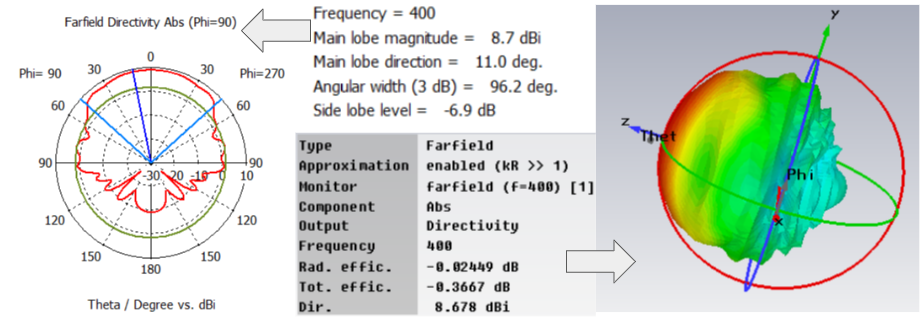}
     \end{subfigure}
     \begin{subfigure}{0.03\textwidth}
     \textbf{D}
     \end{subfigure}
     \begin{subfigure}{0.95\textwidth}
     \centering
     \includegraphics[width=0.8\linewidth,height=\textheight,scale=1,keepaspectratio]{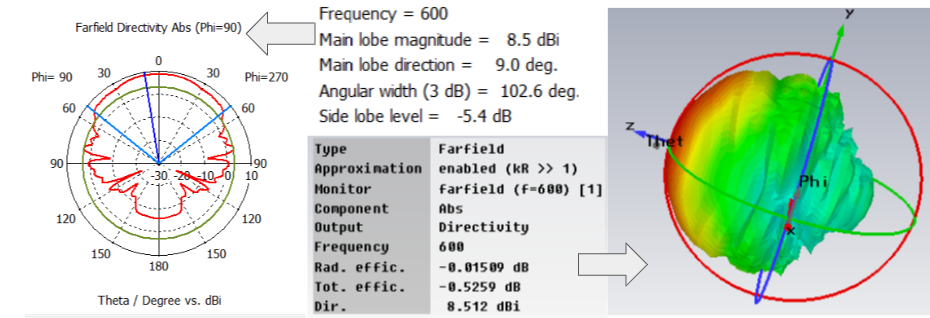}
     \end{subfigure}
     \begin{subfigure}{0.03\textwidth}
      \textbf{E}
      \end{subfigure}
      \begin{subfigure}{0.95\textwidth}
      \centering
      \includegraphics[width=0.8\linewidth,height=\textheight,scale=1,keepaspectratio]{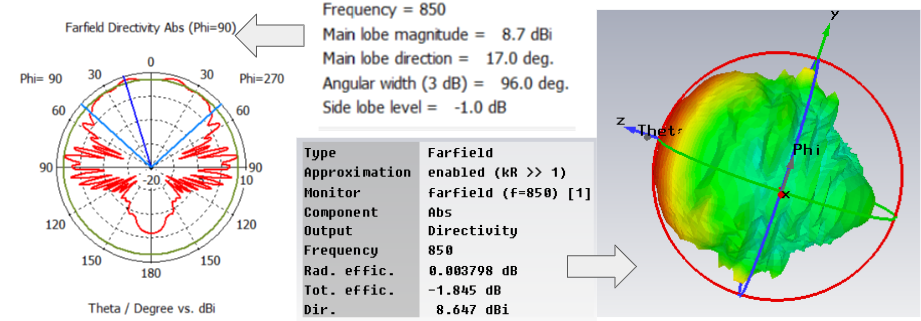}
       \end{subfigure}
    \label{Fig. 4}
    \caption{ H-plane polar plot (left column) and 3D radiation pattern (right column) at different frequencies (in MHz).}
    
\end{figure}
\begin{figure}[t] 
    \centering
    \includegraphics[width=0.9\textwidth,height=\textheight,scale=1,keepaspectratio]{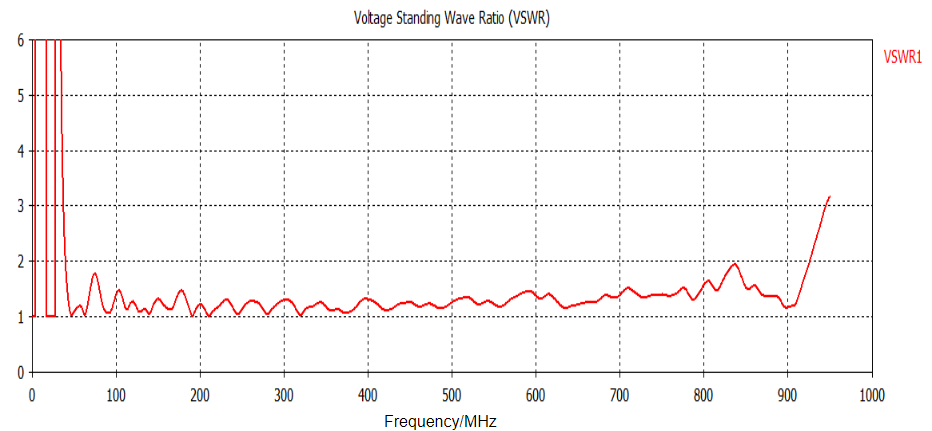}
    \label{Fig. 5}
    \caption{ Plot of Voltage Standing Wave Ratio (VSWR) for the LPDA of CALLISTO system at the USO/PRL.}
\end{figure}

\begin{figure}[t] 
    \centering
    \includegraphics[width=0.9\textwidth,height=\textheight,scale=1,keepaspectratio]{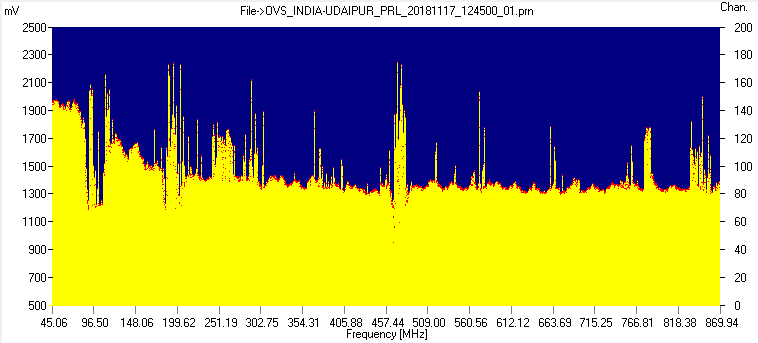}
    \label{Fig. 6}
\caption{ Radio frequency interference (RFI) level measured at the CALLISTO site of USO. We notice the tiny ripple-like patterns on the top of the baseline which represent standing waves in the spectrum.}
\end{figure}

     \begin{figure}[t]
     \centering
   \includegraphics[width=0.5\textwidth,height=\textheight,scale=1,keepaspectratio]{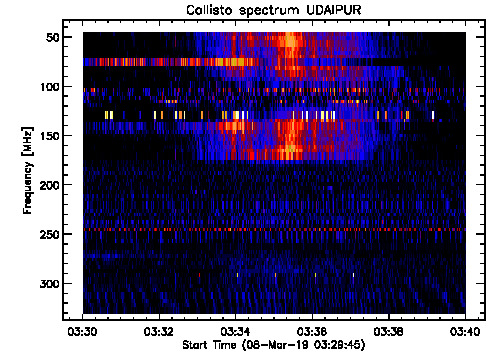}
   \caption{1st light observations of CALLISTO at USO/PRL. The dynamic spectrum shows solar plasma radiation observed on March 8, 2019 during a C-class flare. We note discontinuity in the structure of the radio burst at frequencies ~85-120 MHz which is due to the FM notch. }
      \label{Fig. 7}
  \end{figure}
 \begin{figure}[t]
 \centering
      \includegraphics[width=0.5\textwidth,height=\textheight,scale=1,keepaspectratio]{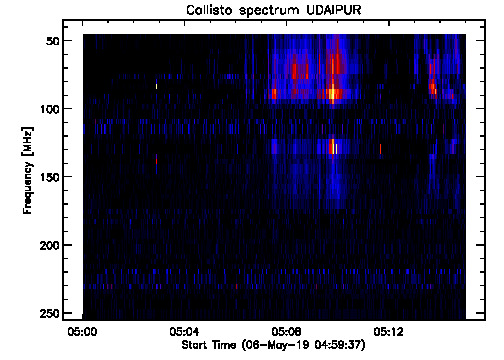}
    \label{Fig. 8}
    \caption{Dynamic radio spectrum showing solar plasma radiation observed on  May 6, 2019.}
\end{figure}

\section{ Summary}
In this article, we present a brief overview of the CALLISTO radio spectrograph which is being operated at the USO/PRL. We provide details of the LPDA which has been developed indigenously. We report some preliminary observations of solar radio bursts taken by our CALLISTO at the USO/PRL and the RFI level at the CALLISTO site is also reported.

\section{Acknowledgement}
 We are  grateful to Dr. Anil Bhardwaj, Director, PRL, Ahmedabad, India for his encouragement and support toward CALLISTO project at the USO/PRL. We also sincerely thank Dr. Yashwant Gupta, Centre Director, NCRA-TIFR, Pune, India for providing technical expertise and facilities to fabricate the LPDA. 

\section{ References}
[1] A. Balanis, Antenna Theory, analysis and design, Constantine Wiley, New York,1989.
\par
[2] Serge Stroobandt,  Michael McCue, Log Periodic Antenna Design Calculator, Available at: https://hamwaves.com/lpda/en/lpda.a4.pdf .
\par
[3] C. Monstein, e-CALLISTO frequency agile radio spectrometer specifications, 
http://www.e-callisto.org/Hardware/eCallistoSpecification.pdf .

\end{document}